\documentclass[apj,numberedappendix]{emulateapj}

\newcommand{\keV}{\mathrm{keV}}							%keV
\newcommand{\MeV}{\mathrm{MeV}}							%MeV

\newcommand{\kmsMpc}{\mathrm{km/s/Mpc}}		  %km/s/Mpc

\newcommand\eqnref[1]{%
  eq.~\ref{eqn:#1}}

\newcommand\figref[1]{%
  \figurename ~\ref{fig:#1}}

\newcommand\secref[1]{%
  Section ~\ref{sec:#1}}

{
\def \today {March 24, 2006}
}

\shorttitle{Sterile neutrinos in the Milky Way}
\shortauthors{Riemer--S{\o}rensen et al.}
\slugcomment{Submitted to ApJL}

\begin{document}

\title{Sterile neutrinos in the Milky Way: \\
Observational constraints}

\author{Signe Riemer-S{\o}rensen$^1$, 
Steen H. Hansen$^2$, and Kristian Pedersen$^1$}

\affil{$^1$ Dark Cosmology Centre, Niels Bohr Institute, University of Copenhagen, Juliane Maries Vej 30, DK-2100 Copenhagen, Denmark\\
$^2$ University of Zurich, Winterthurerstrasse 190,
8057 Zurich, Switzerland}
%\email{}



\begin{abstract}
We consider the possibility of constraining decaying dark matter by
looking out through the Milky Way halo. Specifically we use {\it Chandra} 
blank sky observations to constrain the parameter space of sterile neutrinos.
We find that a broad band in parameter space is still open, leaving
the sterile neutrino as an excellent dark matter candidate.
\end{abstract}




\section{INTRODUCTION}
The particle nature of the dark matter remains a mystery.  This is
contrasted with the firmly established dark matter abundance, which
has been measured using the cosmic microwave background and
large scale structure observations~\citep{spergel,uros}.

A wide range of dark matter particle candidates have been proposed,
including axions, the lightest supersymmetric particle, and sterile
neutrinos. Recently, sterile neutrinos have gained renewed
interest, since theoretical considerations have shown them to be
a natural part of a minimally extended standard model, the
$\nu$MSM. This $\nu$MSM, which is the standard model extended with only 3
sterile neutrinos, manages to explain the masses of the active
neutrinos~\citep{asaka0503065}, the baryon asymmetry of the
universe~\citep{asaka0505013}, and the abundance of dark
matter~\citep{dodelson}.

Arguably, one of the most appealing aspects of sterile neutrinos as 
dark matter candidates is, that they will very likely either be detected or rejected during the next couple of years.  This is because the sterile neutrinos
decay virtually at rest into a photon and an active neutrino
producing a sharp decay line at $E=m_s/2$  where $m_s$ is the mass of the sterile neutrinos~\citep{warm}. This photon line
can now be searched for in cosmic high energy data of the background 
radiation (for recent studies see \citet{mapelli,boyarsky}), 
towards massive structures like galaxy clusters~\citep{tucker,boyarsky06}, 
or towards dark matter structures with a very low fraction of 
baryons~\citep{hansen02}.

The sterile neutrino was originally introduced in order to alleviate
discrepancies between the theory of cold dark matter structure formation and
observations~\citep{dodelson,warm}, which indicated that the dark
matter particle should be warm. Traditionally ``warm'' means that the
mass of the dark matter particle should be in the keV range, in order
to suppress the formation of small scale structures.  While some of these
problems have found other solutions, sterile neutrinos have
found other uses, including an explanation for the pulsar peculiar
velocities~\citep{kusenko,fuller}, synthesising early star
formation~\citep{biermann}, and, as mentioned above, as an explanation
for the masses of the active neutrinos and the baryon
asymmetry~\citep{asaka0503065,asaka0505013}.
Also constraints on doube $\beta$ decays have been derived
\citep{beruzkov}.

We will here follow the idea of~\citet{hansen02} and look for a decay
line towards a region with very little baryons.  Specifically we will
study X-ray ``blank'' regions in the sky with no known X-ray sources.
By doing so we will search for a signal from the dark matter halo of
our own Galaxy. We will see that the signal is very low
for photons with energies of the order 0.1--10~keV, which
allows us to exclude part of the sterile neutrino parameter space.

\begin{figure}[htb]
	\centering
	\includegraphics[angle=270,width=0.49\textwidth]{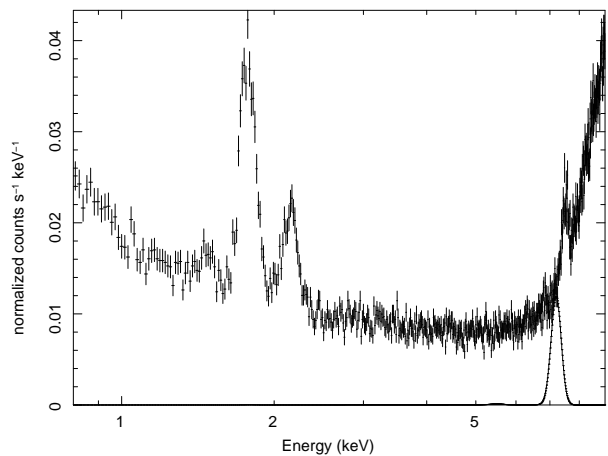}
	\caption{The blank sky ACIS-S3 spectrum -- i.e.\ a view through 
           the Milky Way halo. Also shown is the maximal single gaussian 
           emission line from decaying dark matter at a specific energy as discussed in \secref{analysis}.}
	\label{fig:gaussian}
\end{figure}

\section{X-RAY DATA ANALYSIS} \label{sec:analysis}
We have analysed a 0.3--9~keV spectrum extracted from many combined 
{\it Chandra} ACIS-S3
blank sky observations\footnote{http://cxc.harvard.edu/contrib/maxim}
(shown in \figref{gaussian}) with 
CIAO 3.3\footnote{http://cxc.harvard.edu/ciao}, and fitted spectral models 
with the spectral fitting package Xspec \citep{xspec}.

The line features at 1.74~keV, 2.1--2.2~keV, and 7.48~keV are
Si~$K\alpha$, the Au~$M\alpha\beta$ complex, and Ni~$K\alpha$ respectively,
originating from fluorescence of material in the telescope and focal plane
(i.e.\ they are seen in spectra obtained when ACIS-S3 is stowed and
not pointing at the sky\footnote{http://cxc.harvard.edu/proposer/POG/html/node1.html}). However, a decay line from dark matter could ``hide'' 
under these prominent lines. In this case the decay line should be
redshifted by a factor of $1+z$ where $z$ is the redshift of the dark matter
emitter. Dark matter concentrations located at different distances should thus 
give rise to lines at different observed energies. In order to test for this,
we extracted spectra from ACIS-S3 data of two galaxy clusters, A383 and 
A478, at redshifts, $z=0.1883$ and $0.0881$, respectively.  
We optimized the ratio between expected dark matter signal and emission
from hot intracluster gas by extracting spectra from the cluster outskirts.
The X-ray emission from hot intracluster gas 
is proportional to the gas density squared, whereas the emission of
photons from dark matter is directly proportional to the dark matter 
density. Hence, in the cluster outskirts the gas surface brightness 
falls off with projected cluster-centric distance, $b$ as, 
$S_{gas} \propto b^{-3}$ (assuming a typical $\beta$-model) whereas the dark matter surface brightness is expected to
fall off as $S_{DM} \propto b^{-1}$ (assuming an isothermal sphere). Specifically, we have extracted 
spectra from an outer radial bin, defined by optimizing the ratio of 
matter in the field of view (proportional to the signal from decaying 
dark matter) to the X-ray emission from the intracluster gas (``noise'')
described by a $\beta$-profile. The outer radius was chosen as the radius 
with the optimal ratio (which is very close to the edge of the ACIS-S3 chip) 
and the inner radius was chosen so the signal to noise ratio had a value 
of half the optimal value.
No obvious line features was detected at the corresponding redshifted
energies\footnote{After finisihing this analysis a related paper
appeared~\citep{boyarsky06} where an analysis using exactly this method of
looking at the outer cluster region was made.}.

We then proceeded to firmly constrain the flux from any dark matter
decay line in the full energy interval spanned by the blank sky spectrum.
The blank sky spectrum was well fitted (reduced $\chi ^2 = 1.1$ for 540
d.o.f.) by a composite model consisting of an exponential plus a power
law plus four gaussians for the most prominent lines.
A hypothetical mono-energetic emission line in the spectrum was
represented by a gaussian, centered at the line energy, and with a
width, $\sigma$, given by the instrument spectral resolution: 
$\sigma \approx 0.1$~keV for $0.3\, \keV
\leq E < 6.0$~keV and $\sigma \approx 0.15$~keV for 
$6.0\, \keV \leq E < 8.0$~keV\footnote{http://acis.mit.edu/acis/spect/spect.html}.
(The velocity of the halo dark matter is producing negligible line 
broadening, $v/c \sim 10^{-3}-10^{-4}$).
For each energy in steps of 0.05~keV, we defined a gaussian with the
instrumental width and maximum at the model value of the fit to the 
broad-band spectrum. The flux of this gaussian was then calculated 
providing an upper limit of emission from decaying dark 
matter at the given energy. This is a model independent and very 
conservative analysis, 
since any X-ray emitters within the field of view, and background, 
contributes to the extracted spectrum. If we instead only allowed the 
decay signal from potential decaying dark matter to 
produce a bump above the ``base line'' spectrum, then we could improve the 
bounds by a large factor~\citep{boyarsky}. However, this approach would
depend sensitively on the robustness of modelling background and 
emissions from sources contributing to  the ``base line'' spectrum.

\section{DECAY RATE OF STERILE NEUTRINOS IN THE MILKY WAY HALO}
The amount of dark matter from our Galaxy within the observed field of
view is only a minute fraction of the total mass of the halo.
We use the model for the Milky Way presented in~\citet{klypin}, namely
an NFW dark matter profile with
$M_{vir} = 10^{12} M_\sun$, 
virial radius $r_{vir} = 258$ kpc,
concentration $c=12$,
and solar distance $R_\sun=8$ kpc. We then integrate out through
the Milky Way halo to find the expected signal. Changing the details of
this model does not significantly affrect our results (e.g.\ changing the 
inner density slope between $-1$ and zero, 
or changing the outer slope between $-3$ and $-4$, changes the predicted
signal by less than a factor of 2).
There is also an  uncertainty, possibly as large as a factor of 2,  
arising from using the value of
$M_{vir} = 10^{12} M_\sun$.

The blank sky spectrum was extracted from the full angular opening of 
the {\it Chandra} ACIS-S3 chip which is $8.4 \arcmin$ corresponding to a mass within the field of view of 
$M_{fov}=1.5\times10^{-7}M_{vir}$ at a mean luminosity
distance of $D_L = 35$~kpc, derived from the model mentioned above.

The luminosity from decaying sterile neutrinos is 
$\mathcal{L}=E_{\gamma} N \Gamma _{\gamma}$, where $E_{\gamma}$ is the
energy of the photons, $N$ is the number of sterile neutrinos 
particles in the field of 
view ($N = M_{fov}/m_s$) and $\Gamma _{\gamma}$ is the decay rate. 
The flux at a luminosity distance, $D_L$, is:
\begin{equation}\label{eqn:flux}
F = \frac{\mathcal{L}}{4 \pi D_L^2}=
\frac{E_{\gamma} N \Gamma _{\gamma}}{4\pi D_L^2} \, .
\end{equation}
The observed flux, $F$, at a given energy ($E_{\gamma}=m_s/2$) yields
a conservative upper limit on the flux from decaying sterile neutrinos, so 
\eqnref{flux} can be
rewritten as:
\begin{equation} \label{eqn:maxdecay}
\Gamma _{\gamma} \leq \frac{8 \pi F D_L^2}{M_{fov}}  \, .
%\approx 1.34\times10^{-4}\s^-1 \left(\frac{F}{\ergcms}\right) \left(\frac{D_L}{\Mpc}\right)^2 \left(\frac{M_{fov}}{\Ms}\right)^{-1} \, .
\end{equation}

\begin{figure}[htb]
	\centering
	\includegraphics[width=0.49\textwidth]{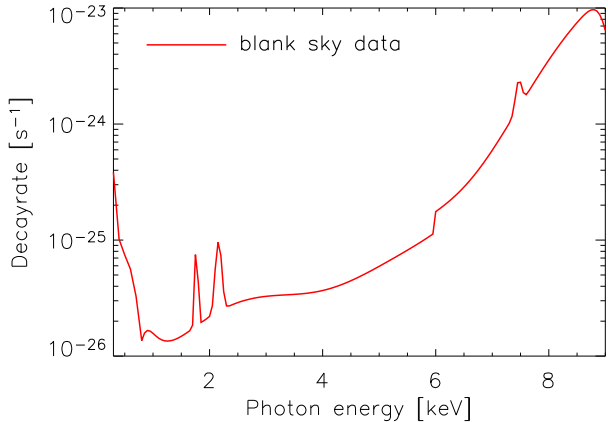}
	\caption{The decay rate as function of photon energy. The
        solid line is the upper limit from the {\it Chandra} blank sky 
	observations viewing out through the Milky Way halo.}
	\label{fig:decayrate}
\end{figure}

From the blank sky flux, an upper limit on the decay rate has been 
derived as function of line energy and plotted in 
\figref{decayrate}.

\section{CONFRONTING THE MODEL WITH DATA}

The neutrinos can decay via various channels, and
the lifetime of the sterile neutrinos is given by \citep{barger:1995,heavy}:
\begin{equation}\label{eqn:tau}
\tau = \frac{1}{\Gamma _{tot}}=\frac{f(m_s)\times 10^{20}}{\left( m_s/\keV \right)^5 \sin^2(2\theta)} \, ,
\end{equation}
$\sin^2(2\theta)$ is the mixing angle with the active neutrino, 
and $f(m_s)$ takes into account the open decay channels so
that for $m_s < 1 \MeV$, where only the neutrino channel is open,
$f(m_s)=0.86$.

The mixing angle is given by \citep[][eqs.~209,~210]{dolgov:2002}:
\begin{eqnarray}
\sin ^2 (2\theta) &\approx& A \left( \frac{\keV}{m_s} \right)^2 
\left( \frac{\Omega
_{DM}}{0.3} \right)\left( \frac{h}{0.65} \right)^2 \nonumber \\
&&  S \left( \frac{g_*(T_{produced})}{10.75} \right)^{3/2}  \, ,
\label{eqn:sin}
\end{eqnarray}
where $g_*(T_{produced})$ is the number of relativistic degrees of
freedom at the temperature at which the sterile neutrinos are
produced. For neutrino masses of $m_s \approx$ keV, the neutrinos are
produced near the QCD phase transition~\citep{dodelson}, making the
distribution somewhat non-thermal~\citep{warm}.  The value of $g_*$
for such neutrinos is very conservatively considered to be
between 10.75 and 20, depending on the details of the QCD phase
transition.  We will here use $g_*(T_{produced})=15$ as a reference value.  
A numerical calculation including the details of the QCD
phase-transition~\citep{kev06} has confirmed that the choice $g_*=15$
gives good agreement with the analytical results presented by
\citet{dolgov:2002}. S is a free parameter taking into account that
additional entropy may be produced after the sterile neutrinos have been
created~\citep{asaka06}. This parameter was suggested to be in the
range between 1 and 100~\citep{asaka06}.  $A$ is a constant depending
on which active neutrino the sterile neutrinos are assumed to mix
with. It takes the values $A_{se}=6.7\times 10^{-8}$ for $\nu _s$
mixing with $\nu _e$ and $A_{s\mu}=4.8\times 10^{-8}$ for $\nu _{\mu,
\tau}$. The sterile neutrinos are assumed to account for all dark matter, 
i.e.\ $\Omega _{DM}=0.30$, and the Hubble parameter
at the present time, $H_0$, given as $h=H_0/(100\kmsMpc)$.

\begin{figure}[htb]
	\centering
	\includegraphics[width=0.49\textwidth]{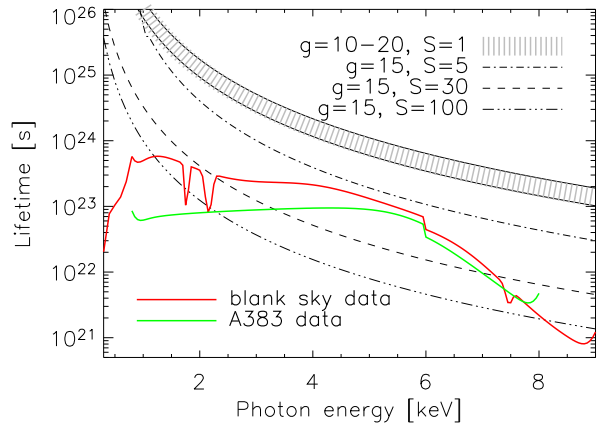}
	\caption{The lifetime constrained from the flux of the {\it Chandra}
	blank sky data (red) and A383 (green). The $\nu$MSM prediction for
	$S=1$ and $g_*=10-20$ (hatched) and several variations of $S$
	and $g_*$ (black) have been overplotted.}
	\label{fig:lifetime}
\end{figure}

\citet{barger:1995} derived the branching ratio for the radiative decay,
$\nu_s \rightarrow \nu_\alpha + \gamma$,
to be $\Gamma _{\gamma}/\Gamma _{tot} = (27\alpha)/(8\pi) \approx 1/128$. 
Now the lifetime can be rewritten as:
\begin{eqnarray}
\tau &=& \left(\frac{E}{\keV} \right)^{-3} \left(\frac{g_*}{15}
\right)^{-3/2} \left(\frac{\Omega _{DM}}{0.3} \right)^{-1}
\left(\frac{h}{0.7} \right)^{-2} \nonumber \\
&& \times \, \frac{f(m_s) 6.5\times10^{18} sec}{A S} \, .
\end{eqnarray}
An upper limit on the decay rate is constrained from \eqnref{maxdecay} and
\begin{equation}
\tau = \frac{1}{\Gamma _{tot}} \approx \frac{1}{128 \Gamma _\gamma} \, .
\end{equation}
The lower limit on the lifetime from the data and the model predictions 
are plotted in \figref{lifetime} for various values of $S$ and $g_*$.

We see from \figref{lifetime} that the blank sky data (solid
red line) is at least one order of magnitude less restrictive than
the simplest sterile model predictions (hatched region). Models with
significant entropy production, $S \approx 30$, are excluded in the
mass range $4.2\, \keV < m < 7 \keV$. For $S>100$ we exclude $2\, \keV
< m < 16 \keV$.

These constraints are more restrictive than the ones obtained
for the Coma cluster periphery \citep{boyarsky06} for masses less than 
approximately $8$ keV.

\section{OTHER CONSTRAINTS}
The simplest models with $S=1$ are bounded from below, $m_s>2\, \keV$,
using Ly$\alpha$ observations
\citep{viel}, and possibly even 
$m_s>14$~keV according to a recent analysis~\citep{seljak}.
These constraints are weakned when allowing for $S>1$~\citep{asaka06},
since additional entropy dilutes the momentum of the sterile neutrinos.

Upper limits on the sterile neutrino mass are derived from the flux of the 
diffuse photon background~\citep{warm}, and are of the order $m_s<15$~keV
for $S=1$ \citep{mapelli,boyarsky}. These bounds are strengthened
for $S>1$, since the sterile neutrinos then will decay faster.

A strict lower bound on the sterile neutrino mass arises since Big Bang
nucleosynthesis only allows approximately $0.3$ additional relativistic 
neutrino species to be populated at $\sim$~MeV temperatures. 
This lower bound is approximately $m_s>50$~eV. The Tremaine-Gunn bound
applied to dwarf speroidals give $m_s>0.5$ keV \citep{tremaine,lin}.

The strong claim of an upper mass limit presented in \cite{tucker},
under-estimated the flux from Virgo cluster by 2 orders of magnitude,
and is unreliable \citep{boyarsky06}.

\begin{figure}[htb]
	\centering
	\includegraphics[width=0.49\textwidth]{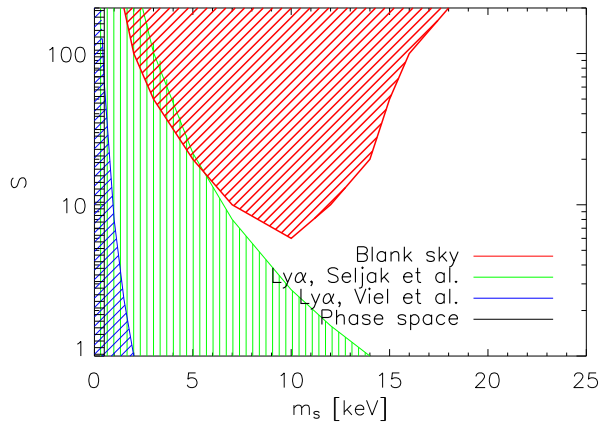}
	\caption{The $m_s-S$ parameter space with constraints from
	the {\it Chandra} blank sky data (red, fat, diagonal, this paper), 
	Ly-$\alpha$ observations
	(blue, diagonal, conservative~\cite{viel}, 
	and green, vertical, ambitious~\cite{seljak}),
	Tremaine-Gunn bound $m>0.5$ keV (black, horizontal). }
	\label{fig:m_S}
\end{figure}

Comparing these existing bounds with our findings, we conclude that a
broad band in neutrino parameter space is still open as shown in
\figref{m_S}. Taking the results of \cite{seljak} into account, then only
masses above approximately $5$ keV are allowed. Ignoring
the results of \cite{seljak}, then another band opens in parameter
space, for small masses and large entropy production.

\section{CONCLUSIONS}
We have used {\it {Chandra}} blank sky observations, i.e.\ viewing out
through the Milky Way halo, to search for a possible decay line from
dark matter particles. No obvious decay line was identified.

The blank sky 0.3--9~keV X-ray flux is used to constrain the parameter 
space of sterile neutrinos, predicted to decay at rest to a detectable photon
and an active neutrino.  We find that the entropy production must
be limited, $S<100$ for the mass range, $2\, \keV < m_s < 16$~keV, 
and even $S<10$ for masses near $m_s=10$ keV.

This leaves a wide allowed space in parameter space, and the sterile
neutrino is therefore still a viable dark matter candidate.  If all
Ly$\alpha$ constraints are considered, then it appears that the
sterile neutrino is either rather massive, or its momenta are diluted
by the entropy production, rendering it a cold dark matter candidate.
If some of the Ly$\alpha$ constraints are relaxed, then there is a
broad allowed band for smaller masses where the sterile neutrino is
still a warm dark matter candidate.

Our analysis is conservative and assumes that the entire blank sky signal 
at a given energy is arising from a decaying neutrino.  By looking for
decay lines above the expected blank sky emission, the results obtained 
here may be improved significantly.

\acknowledgments SHH is pleased to thank Ben Moore for inspiring
discussions.  
We thank Alex Kusenko for useful comments.
The Dark Cosmology Centre is funded by the Danish
National Research Foundation. 
SHH is supported by the Swiss National Foundation.
This research has made use of data
obtained through the High Energy Astrophysics Science Archive Research
Center Online Service, provided by the NASA/Goddard Space Flight
Center.\\

{\em Note added:}\\
After finishing this paper we became aware of an independent analysis,
where the possibility of measuring decaying dark matter particles from our
Galaxy, using the blank sky observations with {\it XMM-Newton}, was considered
\citep{bnrst2006}. The conclusions reached in that paper are qualitatively
similar to ours.\\

Facilities: \facility{Chandra(ACIS)}

\end{document}